\documentclass[runningheads]{llncs}
\usepackage{graphicx}

\usepackage{subcaption}
\usepackage{makecell}
\usepackage{wrapfig}
\usepackage{xcolor}
\usepackage[title]{appendix}

\begin{document}
\title{Jump on the Bandwagon? – Characterizing Bandwagon Phenomenon in Online NBA Fan Communities\thanks{ This is the author version of an article published in SocInfo 2020. The final authenticated version is available online at: https://doi.org/10.1007/978-3-030-60975-7\_30}}
\titlerunning{Characterizing Bandwagon Phenomenon in Online NBA Fan Communities}
%
\author{Yichen Wang \and
Jason Shuo Zhang \and
Xu Han \and
Qin Lv}
\authorrunning{Y. Wang et al.}
%
\institute{Department of Computer Science\\
University of Colorado Boulder, Boulder, CO 80309, USA \\
\email{\{yiwa6864, JasonZhang, xuha2442, Qin.Lv\}@colorado.edu}}
\maketitle              
\begin{abstract}
Understanding user dynamics in online communities has become an active research topic and can provide valuable insights for human behavior analysis and community management. In this work, we investigate the “bandwagon fan” phenomenon, a special case of user dynamics, to provide a large-scale characterization of online fan loyalty in the context of professional sports teams. 
We leverage the existing structure of NBA-related discussion forums on Reddit, investigate the general bandwagon patterns, and trace the behavior of bandwagon fans to capture latent behavioral characteristics. We observe that better teams attract more bandwagon fans, but they do not necessarily come from weak teams. Our analysis of bandwagon fan flow also shows different trends for different teams, as the playoff season progresses. Furthermore, we compare bandwagon users with non-bandwagon users in terms of their activity and language usage. We find that bandwagon users write shorter comments but receive better feedback, and use words that show less attachment to their affiliated teams. Our observations allow for more effective identification of bandwagon users and prediction of users' future bandwagon behavior in a season, as demonstrated by the significant improvement over the baseline method in our evaluation results. 

\end{abstract}

\section{Introduction}
\noindent The proliferation of online social networks has led to the increasing popularity of online communities, which allow like-minded people to connect and  discuss their shared interests, values, and goals without geographic constraints. For instance, in Reddit, a popular online community platform, users can participate in a large number of communities~\cite{Hessel2016-zk,tan2015all}, which are referred to as subreddits and denoted by ``r/" followed by the community's topic, such as r/politics. The plethora of online communities provides a great opportunity to understand human interactions across multiple communities. Some ``loyal" users may commit to one particular community and maintain a stable engagement~\cite{hamilton2017loyalty}, some may jump across several communities~\cite{tan2015all},  while others may change their community affiliation over time.  Such phenomena are fundamental examples of user dynamics in online communities, and understanding how users migrate across communities is an important problem. By analyzing and characterizing these dynamics in terms of users' behavior patterns can provide useful insights for designing better communities, guide community managers to provide better services to improve user engagement, and help community-related stakeholders (e.g., sports teams, celebrities, advertisers) to better promote their business.

The bandwagon phenomenon is a widespread phenomenon in online sports communities. By definition, ``bandwagon fans'' refer to sports fans who start following a sports team only because of its recent success, and this group of fans will be gone immediately after the team performs poorly. Reddit officially introduced the bandwagon mechanism in some sports-related discussion groups (e.g., r/NBA and r/NFL) several years ago, which allows users to change their team affiliation and self-identify themselves as bandwagon fans during playoffs. For instance, during playoffs in NBA season 2016-17, 17.9\% of Cavaliers' fans in r/NBA were bandwagon fans. Our work focuses on examining this specific phenomenon, ``bandwagon fan'', in online NBA fan communities. We leverage the existing structure of NBA-related discussion forums on Reddit to study users' bandwagon behavior in the context of professional sports, a domain that is understudied yet closely connected to people's daily life. We choose online fan groups of professional sports teams as a testbed for the following reasons. First, professional sports play a significant role in modern life and a large population is actively engaged~\cite{guttmann2004ritual,Nielsen2015}. Second, professional sports teams are unambiguously competitive in nature and users affiliated with different teams have clearly different preferences~\cite{zhang2019intergroup}.

\textbf{Present work.} Specifically, using users' posts and comments in r/NBA and individual team subreddits across three  recent NBA seasons, we aim to answer three research questions: (1) What is the general pattern of bandwagon phenomenon and its relationship with team performance? (2) Are there behavioral features that differentiate bandwagon users from non-bandwagon users? and (3) How effective can we identify bandwagon users and predict future bandwagon behavior using these features? Our large-scale study reveals that better teams attract more bandwagon fans and bandwagon fans typically switch to better-performing teams, but not all bandwagon fans come from weak teams. Also, as the playoffs progresses, bandwagon fans from different teams show different team change flow patterns. We also identify clear behavioral differences between bandwagon and non-bandwagon users in terms of their activity and language usage after applying a matching technique, e.g., bandwagon users tend to leave shorter comments but receive better feedback; and they are less attached to their affiliated teams. Using the features we identify, we are able to improve the bandwagon fan classification and prediction results over the bag-of-words baseline method, with 18.9\% and 47.6\% relative improvement, respectively. 

Our work contributes to the following aspects: First, to the best of our knowledge, this is the first large-scale analysis of bandwagon behavior in sports community, which reveals clear behavioral characteristic differences of bandwagon fans compared with non-bandwagon fans. Second, using the observed behavioral characteristics, we can better distinguish bandwagon users and predict future bandwagon behaviors. Third, our work offers new insights for user loyalty research and online community management.

\section{Related Work}
\textbf{User engagement and loyalty in online communities.}
Online community engagement has been a topic of active research~\cite{aldous2019predicting,Mahmud2014-cf,tan2015all,Danescu-Niculescu-Mizil2013-nz,rowe2013changing,zhang2020tale,zhang2017event}. Danescu-Niculescu-Mizil et al.~\cite{Danescu-Niculescu-Mizil2013-nz} build a framework to track users' linguistic change in an intra-community scenario and find they follow a determined two-stage lifecycle: an innovative learning phase and a conservative phase. Tan et al.~\cite{tan2015all} study users' multi-community engagement in Reddit through community trajectory, language usage and feedback received. They find that over time, users span more communities , “jump” more and concentrate less; users' language seems to continuously grow closer to the community’s; frequent posters’ feedback is continually growing more positive; and departing users can be predicted from their initial behavioral features. Loyalty is a fundamental measure of users maintaining or changing affiliation with single or multiple communities. Users' loyalty with a single community is studied via churning in question-answering platform ~\cite{Dror2012-yd,Pudipeddi2014-zc}, where gratitude, popularity related features, and temporal features of posts are shown to be predictive.  Multi-community loyalty in both community and user level is studied by Hamilton et al.~\cite{hamilton2017loyalty}, where loyalty is defined as users making the majority or minority of their posts to a certain forum in a multi-forum site at a specific time. They find loyal communities tend to be smaller, less assertive, less clustered and have denser interaction networks, and user loyalty can be predicted from their first three comments to the community. Different from prior work, our study focuses on the self-defined bandwagon status in sub-communities (different teams) within a large, single community (r/NBA).

\textbf{Bandwagon phenomenon.}
The bandwagon phenomenon is found in many fields such as politics~\cite{McAllister1991-zo}, information diffusion~\cite{nadeau1993new}, sports~\cite{zhang2018we}, and business applications~\cite{sundar2008bandwagon}. In Sundar et al.'s work~\cite{sundar2008bandwagon}, by conducting an experiment using fake products with different star ratings, number of customer reviews, and sales rank on an e-commerce site, they provide the preliminary support for bandwagon effect on users' intention and attitude toward products. Zhu et al.~\cite{Zhu2012-zm} also find that other people’s opinions significantly sway people’s own choices via an experiment in an online recommender system. In Wang et al.'s work~\cite{wang2015bandwagon}, they predict article life cycles in Facebook discussion groups using the bandwagon effect. Voting is another behavior that is related to this phenomenon, where voters may or may not follow their own opinion to make a voting decision but follow the majority~\cite{Kiss2014-ki,Van_der_Meer2016-kr}.
Most of these research efforts conclude at observing the bandwagon phenomenon at the application level. Further study is needed to analyze the specific characteristics, especially in the context of online communities. 

\section{Dataset}

We focus on the professional sports context derived from NBA-related discussion forums (r/NBA and individual team subreddits) on Reddit, an active community-driven platform where users can submit posts and make comments. The r/NBA and team subreddits provide an ideal testbed for observing and understanding bandwagon fans, because a user's team affiliation can be acquired directly through a mechanism known as ``flair''. Flair appears as a short text next to the username in posts and comments (e.g., Lakers $username$). In r/NBA, users can use flairs to indicate support and each user cannot have multiple flairs at the same time. After a specific date (referred to as the \textit{bandwagon date} in the rest of this paper) which is usually shortly before playoffs begin in each season, each user is given the option to change his/her flair to a bandwagon flair of a different team (e.g., Warriors Bandwagon $username$). A user can change the flair as many times as he/she wants. We obtain 0.6M posts and 30M comments as well as their received feedback in NBA-related subreddits from https://pushshift.io~\cite{pushshift2018}, where flair is used to determine the user's bandwagon status. As pointed out by Zhang et al.~\cite{zhang2018we}, offline NBA seasons are reflected in users' behavior in these NBA-related subreddits. As such, we organize our data according to the timeline of NBA seasons and focus on three seasons: 2015-16, 2016-17, and 2017-18. 

\section{General Patterns of Bandwagon Phenomenon}

Our first research question aims to identify the general pattern of bandwagon behavior. We extract all flair changes from team A to team B bandwagon, where A $\neq$ B.
A general user flow network in is shown in Fig. \ref{fig:flows1617}.

\subsubsection{Observation 1: Better teams attract more bandwagon fans, but not all of them come from weak teams.} 
We first investigate which teams bandwagon fans switch to (i.e., target team). Intuitively, we expect the bandwagon fans move to better teams and abandoning the weak ones. Here, we consider the number of bandwagon fans who switched to team B (the target team) and its correlation with either team B's standing (i.e., rank) or the difference between team B' standing and team A (source team)'s standing. Table \ref{tab:corr} shows the correlations computed for each scenario and each season. We can see strong correlations in all but one case. The correlations are negative because lower standing means better performance (e.g., the best team ranks 1), and if team B is better than team A then the standing difference (B's standing - A's standing) should be negative. Based on the results, we can conclude that better teams tend to attract more bandwagon fans.

\begin{table}[t]
\centering
\caption{Correlation results for team standing and standing difference with user count. \textbf{Throughout this paper, the number of stars indicate
p-values:} $^{***}:p < 0.001$, $^{**}:p < 0.01$, $^{*}:p < 0.05$}\label{tab:corr}
\begin{tabular}{|c|c|c|}
\hline
Season &  \makecell{Correlation between team B's \\ standing and user count} & \makecell{Correlation between \\ standing difference and user count} \\
\hline
2015-16 & $-0.675^*$  & $-0.617^{**}$\\
2016-17 & $-0.673^{**}$ & $-0.489^{**}$\\
2017-18 & $-0.789^*$ & 0.188 (not significant)\\
\hline
\end{tabular}
\end{table}

We further study where the bandwagon fans switch from (i.e., the source team A). We examine the correlation of team A's standing with user count. To our surprise, we do not observe any significant correlation for all the three seasons. This indicates that many bandwagon fans also come from strong teams. To better understand this, we calculate each team's bandwagon fan ratio and select the teams with a ratio above the median of all teams. We find that in the three seasons, there are 8, 7, 8 teams respectively which are in the playoffs but have above-median ratio of fans leaving team. For instance, Raptors and Spurs are top 3 in their conference in all three seasons but still have a high percentage of fans leaving. Although bandwagon is only one aspect of loyalty, this result supplements the finding in Zhang et al.'s work on fan loyalty~\cite{zhang2018we}: top teams
\begin{wrapfigure}[24]{R}{0.6\textwidth} 
\vspace{-2.3em}
\hspace*{-.75\columnsep}
\centering
\includegraphics[width=0.6\textwidth]{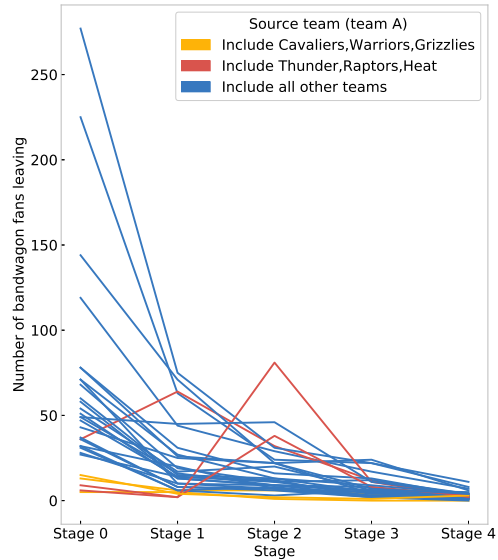}
\caption{\label{fig:trend16}Bandwagon fan flair change trend in season 2015-16.}
\end{wrapfigure}

\noindent tend to have lower fan loyalty, in terms of user retention. As can be seen in Fig.~\ref{fig:flows1617}, not all bandwagon fans come from weak teams and strong teams can also lose a large number of bandwagon fans. 

\subsubsection{Observation 2: Bandwagon flair changes exhibit different stage-wise trends across teams.}
NBA playoffs are elimination tournaments which include 4 rounds: conference first round, conference semi-final, conference final and overall final. To keep our analysis consistent with the temporal structure of NBA playoffs, we divide the period after the \textit{bandwagon date} into 5 stages: the pre-playoffs stage (stage 0) and the other 4 stages corresponding to the 4 rounds in playoffs, to examine the temporal dynamics of user's bandwagon behaviors. As shown in Fig.~\ref{fig:trend16}, we examine the number of bandwagon fans in the 5 stages. We find three types of representative trends: (1) Most teams' bandwagon fans abandon their original teams in stage 0 and there are still fans leaving in later stages but in decreasing numbers. For instance, bottom teams such as  Lakers and Knicks have a large number of vagrant fans, and better teams such as Clippers and Heat also fall into this category due to some vagrant fans. (2) Some teams have very few bandwagon fans in the whole period. Such teams include top ones (e.g., Cavaliers and Warriors) and bottoms ones (e.g., Grizzles). Those teams have a strong fan base. (3) Some teams have a spike during the playoffs, which means most of their bandwagon fans abandon the team in a specific stage and very few leave before or after that stage. Three teams fall into this category in season 2015-16: Thunder, Raptors, and Celtics. These teams have a relatively strong fan base compared with the teams in (1), but some fans lose interest in the middle (e.g., team is eliminated) and choose to support a stronger team. Please note that the teams falling into these three trend categories can vary from season to season depending on each team's performance in a season.

\section{Behavioral Features of Bandwagon Users}
Our second research question aims to identify behavioral features that help differentiate bandwagon fans from non-bandwagon fans. To do this, we apply a matching technique to make the two groups comparable. We first give a formal definition of  \textit{bandwagon users} and identify all these users. Then we match each \textit{bandwagon user} with a \textit{non-bandwagon user} who has a similar activity level. This allows us to directly compare the behavioral features of bandwagon and non-bandwagon fans. 

\subsection{Identify Bandwagon Users for Behavioral Comparison}

To identify bandwagon users within one season, we first define \textit{active} users in r/NBA as those who have at least five activities before the \textit{bandwagon date} and one activity after, where an \textit{activity} refers to either submitting a post or making a comment. 

We view posting/commenting with a team’s flair in r/NBA as an indication of support towards that team. We define an active user as a \textit{fan} of a team during a specific period of time if the user indicates support (flair) only for that team and such support sustains over all activities during that time period. Note that this flair does not contain the word ``bandwagon". If the user only uses flair with ``bandwagon" for a team, we call that user a \textit{bandwagon fan} of that team. 

We further define bandwagon user and non-bandwagon user based on whether a \textit{fan} changes to a \textit{bandwagon fan} of another team. To summarize, we consider the following two groups of users: 
\begin{itemize}
\item {\em Non-bandwagon user}: \textit{Fan} of a team A throughout the whole season.
\item {\em Bandwagon user}: \textit{Fan} of a team A till a time point, and after that point, becomes \textit{bandwagon fan} of team B ( B $\neq$ A) for a period of time, regardless of any bandwagon changes thereafter. 
\end{itemize}
The terms ``fan", ``bandwagon fan", ``bandwagon users" and ``non-bandwagon users" in this section all refer to the definition above.

There are also a number of users who are not in the aforementioned two groups. For instance, one user can be a fan for different teams during the whole season, while others may not be fans of any team. We do not consider those cases in our analysis since they do not link directly to the bandwagon phenomenon that our work focuses on. Statistics of the bandwagon and non-bandwagon users for our analysis are shown in Table~\ref{table:statUsers}.

\begin{table}[t]
\caption{Number of Bandwagon and Non-bandwagon Users}\smallskip
\centering
\smallskip\begin{tabular}{|c|c|c|}
\hline
Season & \#Bandwagon users & \#Non-bandwagon users \\
\hline
2015-16 & 2,562 & 23,165\\
2016-17 & 1,526 & 29,955\\
2017-18 & 1,163 & 36,053\\
\hline
\end{tabular}
\label{table:statUsers}
\end{table}

\subsection{User Matching} To make a fair comparison between the two user groups, we need to rule out the influence of activity level. Specifically, for each bandwagon user who is a fan of team A at the beginning, we find a matching non-bandwagon user who has a similar activity level and supports the same team. As a result, we have 2562, 1526, and 1163 user pairs after user matching in the three seasons, respectively.

To evaluate the result of our matching procedure, we check distributional differences in terms of the number of activities between the treatment group (bandwagon users) and the control group (non-bandwagon users). We compare their empirical cumulative distributions before and after matching, using the Mann-Whitney U test~\cite{mann1947test}.  Prior to matching, the p-value for this feature is very close to 0.0, indicating significant difference. After matching, we find no difference between the treatment group and the matched control group at the 5\% significance level in all three seasons ($p$ = 0.485, 0.489, 0.49 for the three seasons), indicating that the data is balanced in terms of activity level after user matching. Fig.~\ref{fig:uTest} in the Appendix shows the CDF plots before and after matching.
\noindent Using the matched user groups, we then characterize how bandwagon users behave differently in terms of their posting/commenting activities and language usage. For consistency, we only consider user activities that occur before the \textit{bandwagon date}. Since users cannot change their flairs to bandwagon flairs prior to that date, the differential features we observe can also be used for predicting future bandwagon behavior in playoffs.

\subsection{Activity Features}

\subsubsection{Observation 1: Bandwagon users are less active than non-bandwagon users in individual team subreddit.}

We compare users' activity level in the same individual team subreddit after matching in r/NBA, and find that bandwagon users have fewer activities (i.e., more silent) than non-bandwagon users in terms of posting/commenting count: (19.91 vs. 29.40$^*$, 23.55 vs. 35.63$^*$, 23.51 vs. 32.4$^*$ for the three seasons, respectively). This is reasonable since individual team subreddits tend to attract fans who are more loyal/dedicated to their teams and participate more actively in their subcommunities.

\subsubsection{Observation 2: Bandwagon users write shorter comments but receive better feedback in r/NBA.} Here, we compute a score for each comment (\#upvotes - \#downvotes) as a measure of received feedback. The higher the score, the better the feedback it receives. Our results show that bandwagon users write significantly shorter comments (18.71 vs. 20.75$^{**}$, 18.34 vs. 20.30$^{**}$, 18.26 vs. 19.55$^{**}$ for the three seasons, respectively), but receive better feedback (11.08 vs. 9.88$^{*}$, 13.11 vs 11.80$^{*}$, 16.64 vs 16.32$^{*}$ for the three seasons, respectively). This observation is consistent with the general pattern that users who ``wander around'' across diverse communities are more likely to receive better feedback~\cite{tan2015all}.

\subsection{Language Usage Features}
Although we use both posting and commenting as indicators of user activity level, we focus more on comments in language usage analysis since posts are more about news and game reports, and less reflective of users' personal characteristics.
\subsubsection{Observation 3: Bandwagon users talk less about specific players and teams.}
To analyze the content of users' comments, we first conduct preprocessing steps including lowercasing words and removing stop words and hyperlinks from comments. After that, Latent Dirichlet Allocation (LDA)~\cite{blei2003latent}, a widely-used topic modeling method, is applied to extract keywords and topics. In our case the perplexity score drops significantly when the number of topics increases from 5 to 10, but remains stable afterwards (from 10 to 30). Therefore, we use 10 as our topic number in this analysis.  

We compare the topics of bandwagon comments and that of  non-bandwagon ones. While they share many similar topics (e.g., game strategy related words: defense, offense; emotional expressions related words: shit, lol), there still exist some differences. Non-bandwagon users talk more about specific teams/players  (e.g., Harden, spursgame), even when they are talking about similar topics as bandwagon users. It shows that bandwagon users appear to be less concerned about the details of teams/players, indicating a relatively indifferent attitude towards their affiliated teams.

\subsubsection{Observation 4: Bandwagon users are less dedicated to discussions in terms of word usage.}
Inspired by Hamilton et al.~\cite{hamilton2017loyalty} and previous observations, we examine the two groups' word usage. To capture the esoteric content and users' attachment to the teams, we calculate the proportion or summary variable of different types of words in their comments using LIWC word categories~\cite{pennebaker2015development}, a well known set of word categories that were created to capture people’s social and psychological states. We find significantly less word use of bandwagon comments in five word categories: clout~\cite{kacewicz2014pronoun} (high clout value suggests that the author is texting from the perspective of high expertise and is confident), social process (words that suggest human interaction), cognitive process (words that suggest cognitive activity), drives (an overarching dimension that captures the words that represent the needs, motives and drives including achievement, power, reward, etc.), and future focus (future tense verbs and references to future events/times). The results are shown in Table \ref{table:wordUsage}. Please note that the results show the average value of bandwagon users versus that of non-bandwagon users. All these five lexicon categories show that non-bandwagon users have a closer attachment to their affiliated teams and a more proactive attitude towards discussions. 

\begin{table}[t]
\caption{LIWC Word Categories Analysis Results}\smallskip
\centering
\smallskip\begin{tabular}{|l|l|l|l|}
\hline
Word category & Season 2015-16&Season 2016-17&Season 2017-18 \\
\hline
clout & 52.36 vs. 53.76$^{***}$ &52.24 vs. 53.07$^{**}$& 52.13 vs. 53.39$^{***}$\\
social process & 9.26 vs. 9.42$^{*}$ &9.45 vs. 9.70$^{*}$&9.48 vs. 9.74$^{*}$\\
cognitive process & 10.52 vs. 10.86$^{***}$ &10.52 vs. 10.90$^{**}$ & 10.40 vs. 10.64 ($p < 0.1$)\\
drives & 7.44 vs. 7.58$^{*}$ &7.59 vs. 7.77$^{*}$ &7.41 vs. 7.59 ($p < 0.1$)\\
future focus & 1.09 vs. 1.13$^{*}$ & 1.10 vs. 1.18$^{**}$&1.12 vs. 1.19$^{*}$\\
\hline
\end{tabular}
\label{table:wordUsage}
\end{table}

\section{Bandwagon User Classification and Prediction}

To demonstrate the differential power of the behavioral features we have identified in the previous section, we formulate a classification task and a prediction task to investigate how the activity and language usage features can be used for identifying bandwagon users and inferring future bandwagon behavior. 

\subsection{Experimental Setup}
\subsubsection{Tasks.} Our first task is a classification task that aims to distinguish between bandwagon and non-bandwagon users. We take all users (after matching) across the three seasons, and randomly select 80\% of the users' data as the training set and the remaining 20\% as the testing set.

Our second task is to predict whether a user will become a bandwagon user (i.e., change his/her flair to a bandwagon flair) during a season, based on his/her behavior before the bandwagon date in that season. We take users' data in season 2015-16 and 2016-17 to train our model, and apply it to the data in season 2017-18 to predict if a user will jump on the bandwagon in that season.

\smallskip
\noindent\textbf{Features.} Based on previous observations, we extract two types of features to train our classification and prediction models.
\begin{itemize}
\item Activity features: This set of features includes average comment length and average comment feedback score as discussed in the previous section.
\item Language features: This set of features includes the average summary variable of clout~\cite{kacewicz2014pronoun},  average word proportion in terms of social process, cognitive process, drives, and future focus, as discussed in the previous section.
\end{itemize}

We use Bag-of-words (BOW) features as a strong baseline since BOW not only effectively captures the content of users' comments~\cite{harris1954distributional}, but also requires no pre-observational study.  Please note that all the aforementioned features are extracted from the period between season beginning and \textit{bandwagon date}. 

\smallskip
\textbf{Evaluation procedure.} As the analyses before, we conduct the same user matching process in the first place. We label bandwagon users as positive examples. To evaluate the effectiveness of the classification and prediction tasks using the behavioral features we have identified, we deploy a standard $l2$-regularized logistic regression classifier, and use grid search to find the best parameters. All the results in the classification task are derived after 5-fold cross-validation. 
We consider both precision and recall as the evaluation metrics.

\subsection{Results}

\begin{figure}
\centering
\vspace{-1em} 
\includegraphics[width=0.75\textwidth]{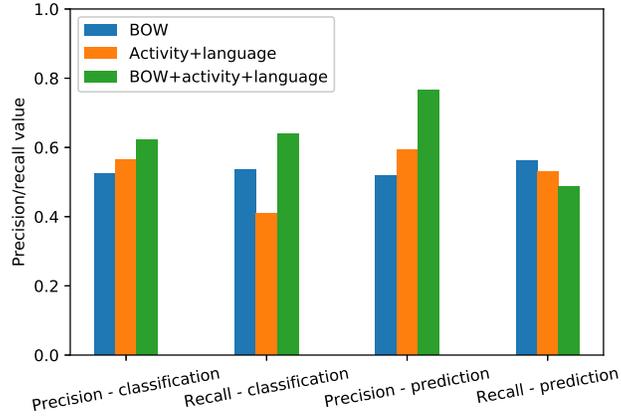} 

\caption{Classification and prediction performance using  different feature sets.}
\label{fig:pred}
\end{figure}

Fig.~\ref{fig:pred} summarizes the  classification and prediction performance when using different feature sets. 
As shown in the figure, the two sets of behavioral features that we identify (activity and language) 
improve precision for both tasks. When combined with BOW features, there is further improvement (18.9\% for classification and 47.6\% on prediction, as compared with the baseline result). These results indicate that the  activity and language features we identify are good complements for BOW text representations to reduce false positive.  

However, we do notice that our activity and language features do not work as well on improving recall in both tasks. Although they improve the classification recall when combined with BOW features, the recall is lower than the BOW only scenario. One possible reason is that some non-bandwagon users actually behave as bandwagon ones but fail to report their bandwagon flairs on Reddit. We find  some non-bandwagon users who start to follow another stronger team's games and news after their original affiliated teams are eliminated, but do not change their flair, especially for users in season 2017-18. These ``fake" non-bandwagon users can confuse our model, resulting in missed bandwagon detection. The recall performance of using all features for the classification task has been improved because we include all three seasons' users in our training set and the combined features can catch some good indicators of ``fake'' non-bandwagon users for each season, while for the prediction task, the training data does not contain any users' information in the 3rd season, 2017-18.

\section{Concluding Discussion}
In this work, we have analyzed the bandwagon phenomenon (a common case of user dynamics) using NBA-related subreddits data from Reddit. We find that better teams attract more bandwagoners, but bandwagoners do not necessarily come from weak teams. Most teams' vagrant fans leave their teams and jump on another team's bandwagon at the beginning, while some teams have a relatively stronger fan base and their bandwagon fans leave when the teams are eliminated. In the comparison after user matching, we find that bandwagon users write shorter comments but receive better feedback, and use words that show less attachment to their affiliated teams. These features can effectively help classify bandwagon users and predict users' future bandwagon behavior.

\subsubsection{Implications for user loyalty.} 
Our results show that loyalty plays an important role in online communities. Bandwagoners have clear behavioral differences from non-bandwagoners. It is crucial for community managers to identify loyal and vagrant fans with the goals of maintaining and growing their user base. To this end, our classification and prediction models show the feasibility of automating these identification processes, and demonstrate the great potential of incorporating such capabilities as a more standard pipeline.

Our work also complements research on user and multi-community engagement, and offers insights on how users behave across sub-communities. The bandwagon phenomenon in our study is about user's preference change in sub-communities, where users share the common interest (basketball), but have different preferences towards the teams. 

In addition, we find that bandwagoning in r/NBA is different from general loyalty. We notice that around 80\% of the bandwagoners in seasons 2015-16 and 2016-17 change their flairs back to their original teams in the following season, which means that bandwagoning is a ``temporary" non-loyal behavior for most users. Thus, one possible future direction is to investigate what factors account for their choice of bandwagoning and willingness to stay with the new team.

\subsubsection{Implications for sports community management.} 
Our findings on bandwagon users' characteristics can be useful for sports team management.
Firstly, our observations reveal that bandwagon fans are not necessarily from weak teams, which suggest that some higher-ranked teams also need to pay close attention to maintaining their fan base. As mentioned earlier, Spurs is a good example. Furthermore, since bandwagon users tend to move ``up" to higher-ranked teams, a strategy to gain some temporary support for the strong-but-not-top teams is to attract more ``travelers" during the playoffs. For example, during the 2016 western semi-final between Thunder and Warriors, Thunder acted as a challenger and highlighted their two star players Kevin Durant and Russell Westbrook. These actions brought them lots of fans from other teams, especially from the teams that were defeated by Warriors.

Secondly, it is important to keep fans engaged in online discussion during off-season, especially when certain fans' affiliated teams are eliminated. Prior work has shown that incorporating group identity can help strengthen member attachment in online communities~\cite{ren2012building}. Our results show that this bandwagon mechanism, i.e., allowing users to switch team affiliation, does have some effective impact on not only encouraging some weak teams' fans to change their flairs and participate in other teams' discussion, but also encouraging certain strong teams' fans to go ``up" to a better team when their teams are eliminated during playoffs. 

\subsubsection{Limitations and future work.} One key limitation of our work is the representativeness of our dataset. Although Goldschein~\cite{Goldschein2016} suggests that /r/NBA is now playing an important role among fans, the NBA fan communities
on Reddit may not be representative of the whole communities. Another limitation is that the bandwagon identity requires users' self-identification. As discussed earlier, the recall is low because there are some ``fake" loyal users who do not use bandwagon flairs but act the same as \textit{bandwagon users}. We also notice that the number of \textit{bandwagon users} is decreasing, which means that fewer users are ``serious" about using this bandwagon flair mechanism. Future directions to address this include designing better strategies in online communities to promote user behavior diversity, and designing better metrics and algorithms to identify real vagrant fans.

Another important question to ask is why users choose to bandwagon. In our analysis we find some fans jump on the bandwagon because their original teams are eliminated and they turn to another team just to have something to watch. Another finding is that some fans jump to teams which are opponents of their ``enemy'' team, i.e., ``Enemy of my enemy is my friend". Answering this question will help provide a fundamental explanation to the bandwagon behaviors. 

In addition to online community, the bandwagon effect also plays an important role in information diffusion~\cite{nadeau1993new}, and impacts the propagation of fake news ~\cite{shao2018anatomy}, where the popularity of news allows users to bypass the responsibility of verifying information. One future direction is to investigate how bandwagoning affects users and helps fake news spread, and how to identify the impacted users.

%
%
%
\bibliographystyle{splncs04}
\bibliography{ref}

\begin{thebibliography}{10}
\providecommand{\url}[1]{\texttt{#1}}
\providecommand{\urlprefix}{URL }
\providecommand{\doi}[1]{https://doi.org/#1}

\bibitem{aldous2019predicting}
Aldous, K.K., An, J., Jansen, B.J.: Predicting audience engagement across
  social media platforms in the news domain. In: SocInfo. pp. 173--187.
  Springer (2019)

\bibitem{pushshift2018}
Baumgartner, J.: Reddit dataset. \url{https://files.pushshift.io/reddit/}
  (2018)

\bibitem{blei2003latent}
Blei, D.M., Ng, A.Y., Jordan, M.I.: Latent dirichlet allocation. JMLR
  \textbf{3}(Jan),  993--1022 (2003)

\bibitem{Danescu-Niculescu-Mizil2013-nz}
Danescu-Niculescu-Mizil, C., West, R., Jurafsky, D., Leskovec, J., Potts, C.:
  No country for old members: user lifecycle and linguistic change in online
  communities. In: WWW. pp. 307--318. WWW '13, ACM, New York, NY, USA (May
  2013)

\bibitem{Dror2012-yd}
Dror, G., Pelleg, D., Rokhlenko, O., Szpektor, I.: Churn prediction in new
  users of yahoo! answers. In: WWW. pp. 829--834. WWW '12 Companion,
  Association for Computing Machinery, New York, NY, USA (Apr 2012)

\bibitem{Goldschein2016}
Goldschein, E.: It’s time to give /r/nba the respect it deserves.
  \url{https://www.sportsgrid.com/as-seen-on-tv/media/its-time-to-give-rnba-the-respect-it-deserves}
  (2015)

\bibitem{guttmann2004ritual}
Guttmann, A.: From ritual to record: The nature of modern sports. Columbia
  University Press (2004)

\bibitem{hamilton2017loyalty}
Hamilton, W.L., Zhang, J., Danescu-Niculescu-Mizil, C., Jurafsky, D., Leskovec,
  J.: Loyalty in online communities. In: ICWSM (2017)

\bibitem{harris1954distributional}
Harris, Z.S.: Distributional structure. Word  \textbf{10}(2-3),  146--162
  (1954)

\bibitem{Hessel2016-zk}
Hessel, J., Tan, C., Lee, L.: Science, {AskScience}, and {BadScience}: On the
  coexistence of highly related communities. In: ICWSM (Mar 2016)

\bibitem{kacewicz2014pronoun}
Kacewicz, E., Pennebaker, J.W., Davis, M., Jeon, M., Graesser, A.C.: Pronoun
  use reflects standings in social hierarchies. Journal of Language and Social
  Psychology  \textbf{33}(2),  125--143 (2014)

\bibitem{Kiss2014-ki}
Kiss, {\'A}., Simonovits, G.: Identifying the bandwagon effect in two-round
  elections. Public Choice  \textbf{160}(3),  327--344 (Sep 2014)

\bibitem{Mahmud2014-cf}
Mahmud, J., Chen, J., Nichols, J.: Why are you more engaged? predicting social
  engagement from word use. arXiv preprint arXiv:1402.6690  (2014)

\bibitem{mann1947test}
Mann, H.B., Whitney, D.R.: On a test of whether one of two random variables is
  stochastically larger than the other. The annals of mathematical statistics
  pp. 50--60 (1947)

\bibitem{McAllister1991-zo}
McAllister, I., Studlar, D.T.: Bandwagon, underdog, or projection? opinion
  polls and electoral choice in britain, 1979-1987. J. Polit.  \textbf{53}(3),
  720--741 (Aug 1991)

\bibitem{Van_der_Meer2016-kr}
van~der Meer, T.W.G., Hakhverdian, A., Aaldering, L.: Off the fence, onto the
  bandwagon? a {Large-Scale} survey experiment on effect of {Real-Life} poll
  outcomes on subsequent vote intentions. Int J Public Opin Res
  \textbf{28}(1),  46--72 (Mar 2016)

\bibitem{nadeau1993new}
Nadeau, R., Cloutier, E., Guay, J.H.: New evidence about the existence of a
  bandwagon effect in the opinion formation process. International Political
  Science Review  \textbf{14}(2),  203--213 (1993)

\bibitem{Nielsen2015}
Nielsen.com: The year in sports media report: 2015.
  \url{https://www.nielsen.com/us/en/insights/reports/2016/the-year-in-sports-media-report-2015.html}
  (2016)

\bibitem{pennebaker2015development}
Pennebaker, J.W., Boyd, R.L., Jordan, K., Blackburn, K.: The development and
  psychometric properties of liwc2015. Tech. rep. (2015)

\bibitem{Pudipeddi2014-zc}
Pudipeddi, J.S., Akoglu, L., Tong, H.: User churn in focused question answering
  sites: characterizations and prediction. In: WWW. pp. 469--474. WWW '14
  Companion, Association for Computing Machinery, New York, NY, USA (Apr 2014)

\bibitem{ren2012building}
Ren, Y., Harper, F.M., Drenner, S., Terveen, L., Kiesler, S., Riedl, J., Kraut,
  R.E.: Building member attachment in online communities: Applying theories of
  group identity and interpersonal bonds. Mis Quarterly pp. 841--864 (2012)

\bibitem{rowe2013changing}
Rowe, M.: Changing with time: Modelling and detecting user lifecycle periods in
  online community platforms. In: SocInfo. pp. 30--39. Springer (2013)

\bibitem{shao2018anatomy}
Shao, C., Hui, P.M., Wang, L., Jiang, X., Flammini, A., Menczer, F.,
  Ciampaglia, G.L.: Anatomy of an online misinformation network. PloS one
  \textbf{13}(4),  e0196087 (2018)

\bibitem{sundar2008bandwagon}
Sundar, S.S., Oeldorf-Hirsch, A., Xu, Q.: The bandwagon effect of collaborative
  filtering technology. In: CHI'08 extended abstracts on Human factors in
  computing systems. pp. 3453--3458. ACM (2008)

\bibitem{tan2015all}
Tan, C., Lee, L.: All who wander: On the prevalence and characteristics of
  multi-community engagement. In: Proceedings of the 24th International
  Conference on World Wide Web. pp. 1056--1066 (2015)

\bibitem{wang2015bandwagon}
Wang, K.C., Lai, C.M., Wang, T., Wu, S.F.: Bandwagon effect in facebook
  discussion groups. In: Proceedings of the ASE BigData \& SocialInformatics
  2015. p.~17. ACM (2015)

\bibitem{zhang2020tale}
Zhang, J.S., Keegan, B.C., Lv, Q., Tan, C.: A tale of two communities:
  Characterizing reddit response to covid-19 through/r/china\_flu
  and/r/coronavirus. arXiv preprint arXiv:2006.04816  (2020)

\bibitem{zhang2017event}
Zhang, J.S., Lv, Q.: Event organization 101: Understanding latent factors of
  event popularity. In: ICWSM (2017)

\bibitem{zhang2018we}
Zhang, J.S., Tan, C., Lv, Q.: This is why we play: Characterizing online fan
  communities of the nba teams. Proceedings of the ACM on Human-Computer
  Interaction  \textbf{2}(CSCW), ~197 (2018)

\bibitem{zhang2019intergroup}
Zhang, J.S., Tan, C., Lv, Q.: Intergroup contact in the wild: Characterizing
  language differences between intergroup and single-group members in
  nba-related discussion forums. Proceedings of the ACM on Human-Computer
  Interaction  \textbf{3}(CSCW),  1--35 (2019)

\bibitem{Zhu2012-zm}
Zhu, H., Huberman, B., Luon, Y.: To switch or not to switch: understanding
  social influence in online choices. In: Proceedings of the SIGCHI Conference
  on Human Factors in Computing Systems. pp. 2257--2266 (2012)

\end{thebibliography}
\vspace*{\fill}
\appendix
\section{Appendix}
\begin{figure}[ht] 
    \begin{subfigure}[b]{.5\linewidth}
      \includegraphics[width=.99\textwidth]{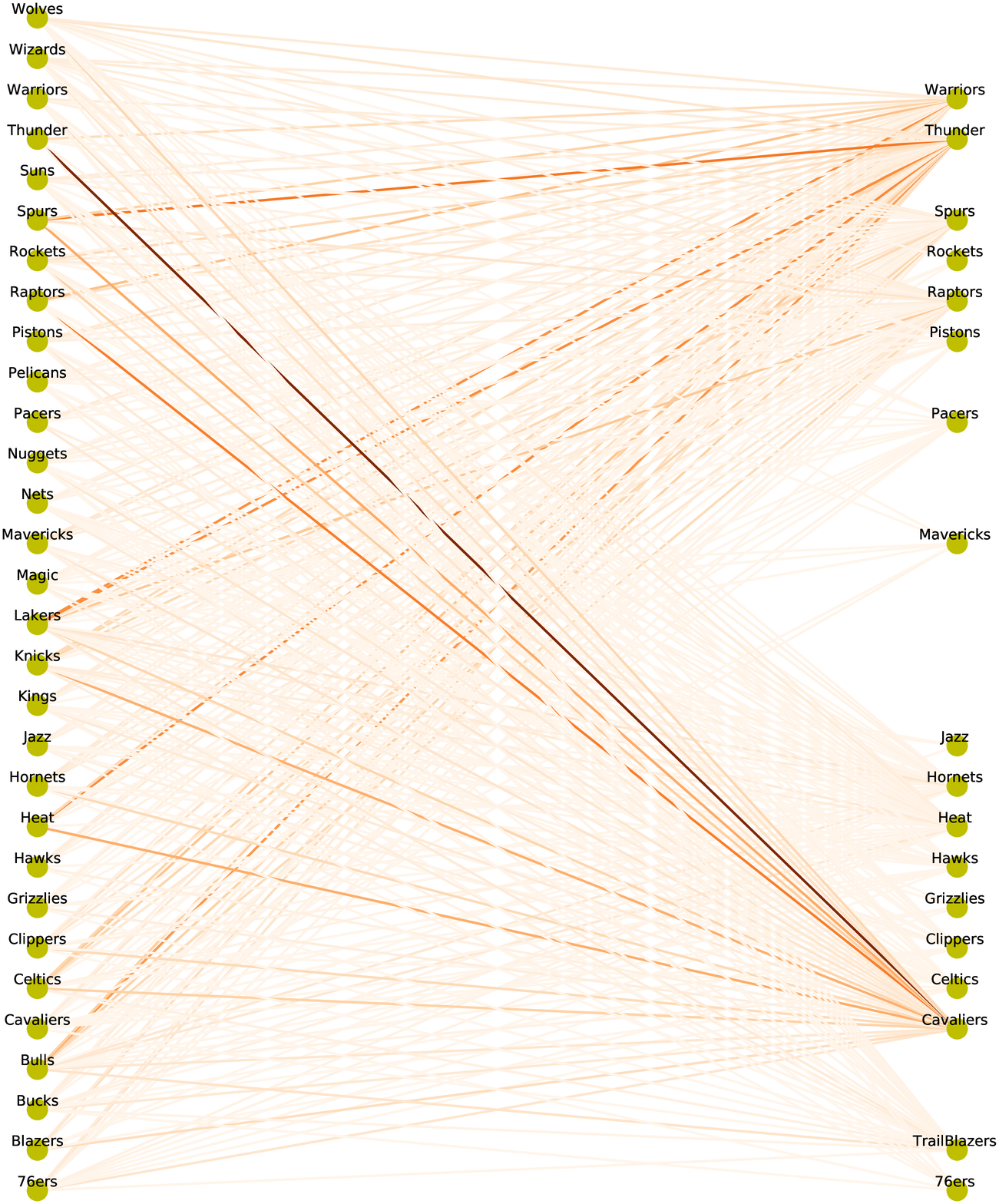}
      \caption{Bandwagon fan flair change flow in season 2015-16}
    \end{subfigure}
    \begin{subfigure}[b]{.5\linewidth}
      \includegraphics[width=.99\textwidth]{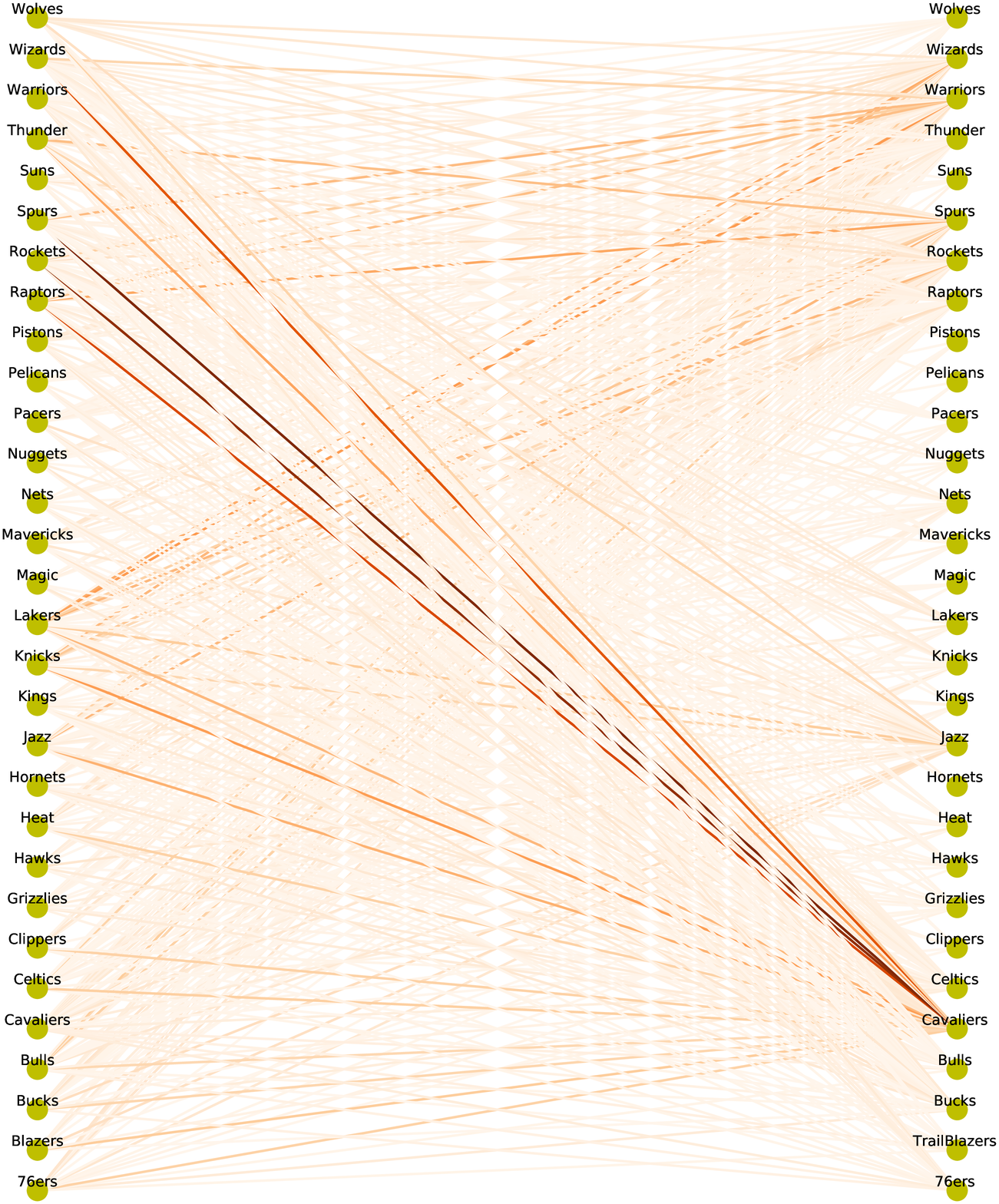}
      \caption{Bandwagon fan flair change flow in season 2016-17}
    \end{subfigure}
\caption{General bandwagon fan flair change flow in different stages in seasons 2015-16 and 2016-17: from source team (team A, left nodes) to target team (team B, right nodes). }
\label{fig:flows1617}
\end{figure}

\begin{figure}[ht]
    \begin{subfigure}[b]{.333\linewidth}
      \includegraphics[width=.99\textwidth]{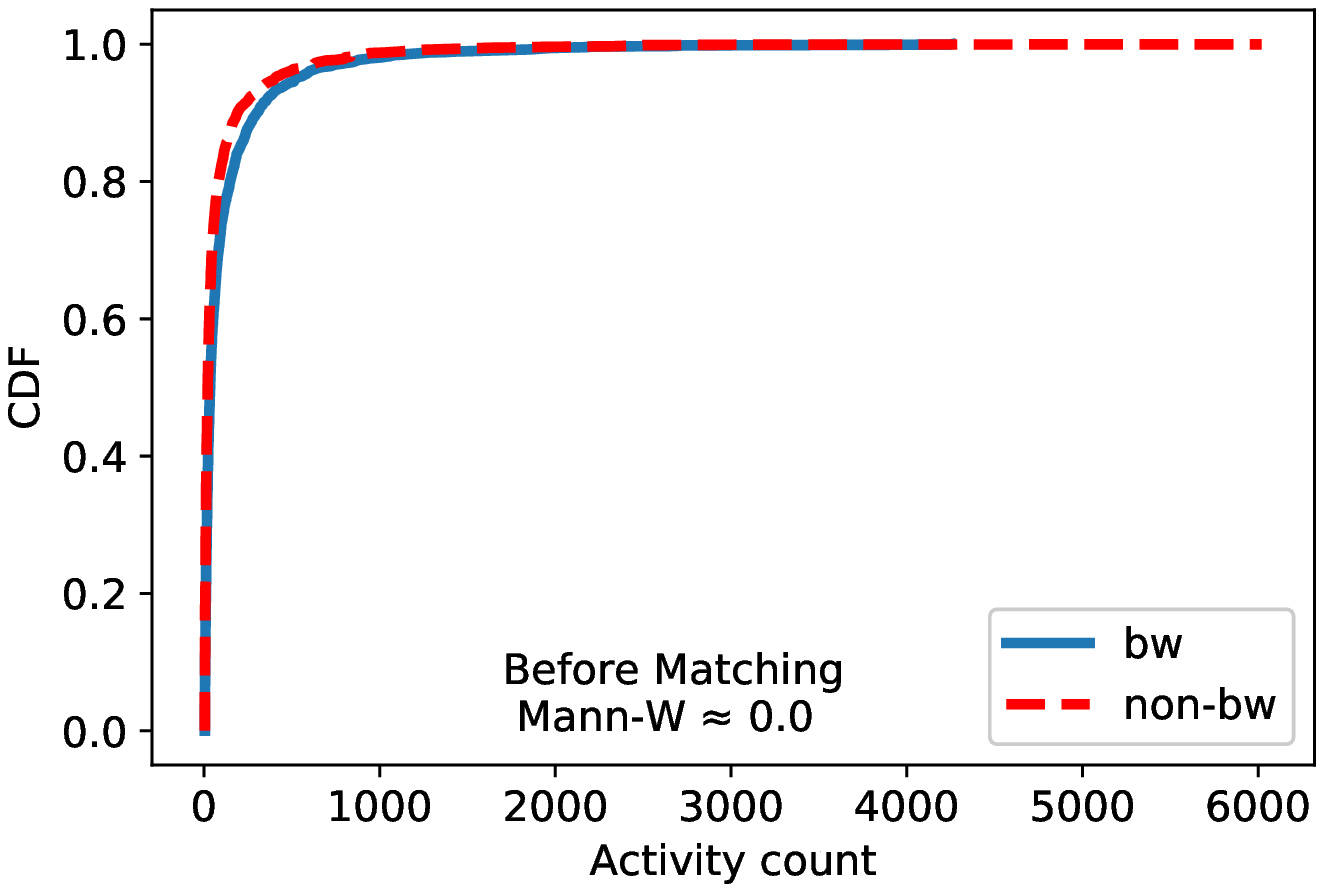}
      \caption{Before, 2015-16}
    \end{subfigure}\hfil
    \begin{subfigure}[b]{.333\linewidth}
      \includegraphics[width=.99\textwidth]{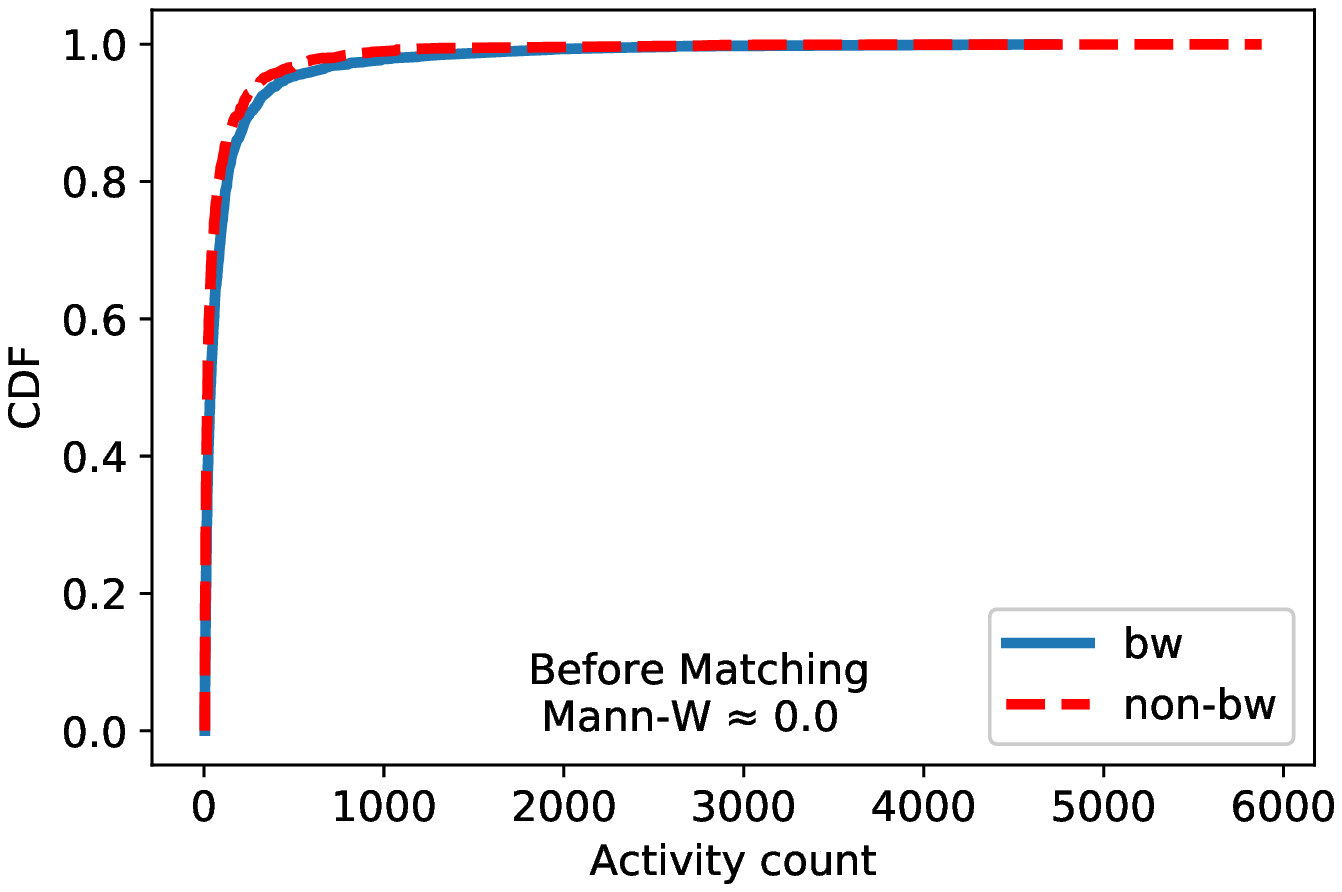}
      \caption{Before, 2016-17}
    \end{subfigure}\hfil
    \begin{subfigure}[b]{.333\linewidth}
      \includegraphics[width=.99\textwidth]{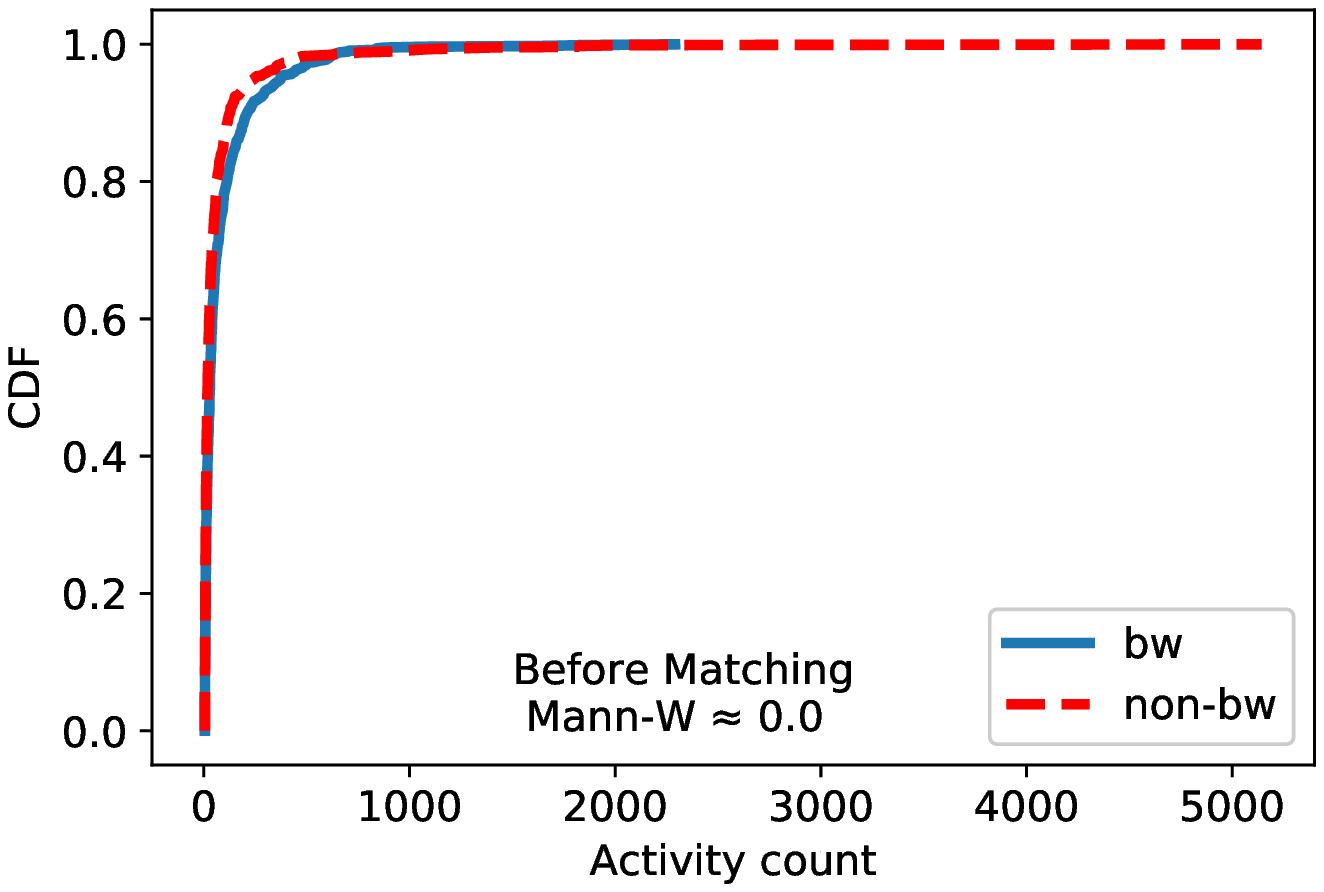}
      \caption{Before, 2017-18}
    \end{subfigure}
    
    \medskip
    \begin{subfigure}[b]{.333\linewidth}
      \includegraphics[width=.99\textwidth]{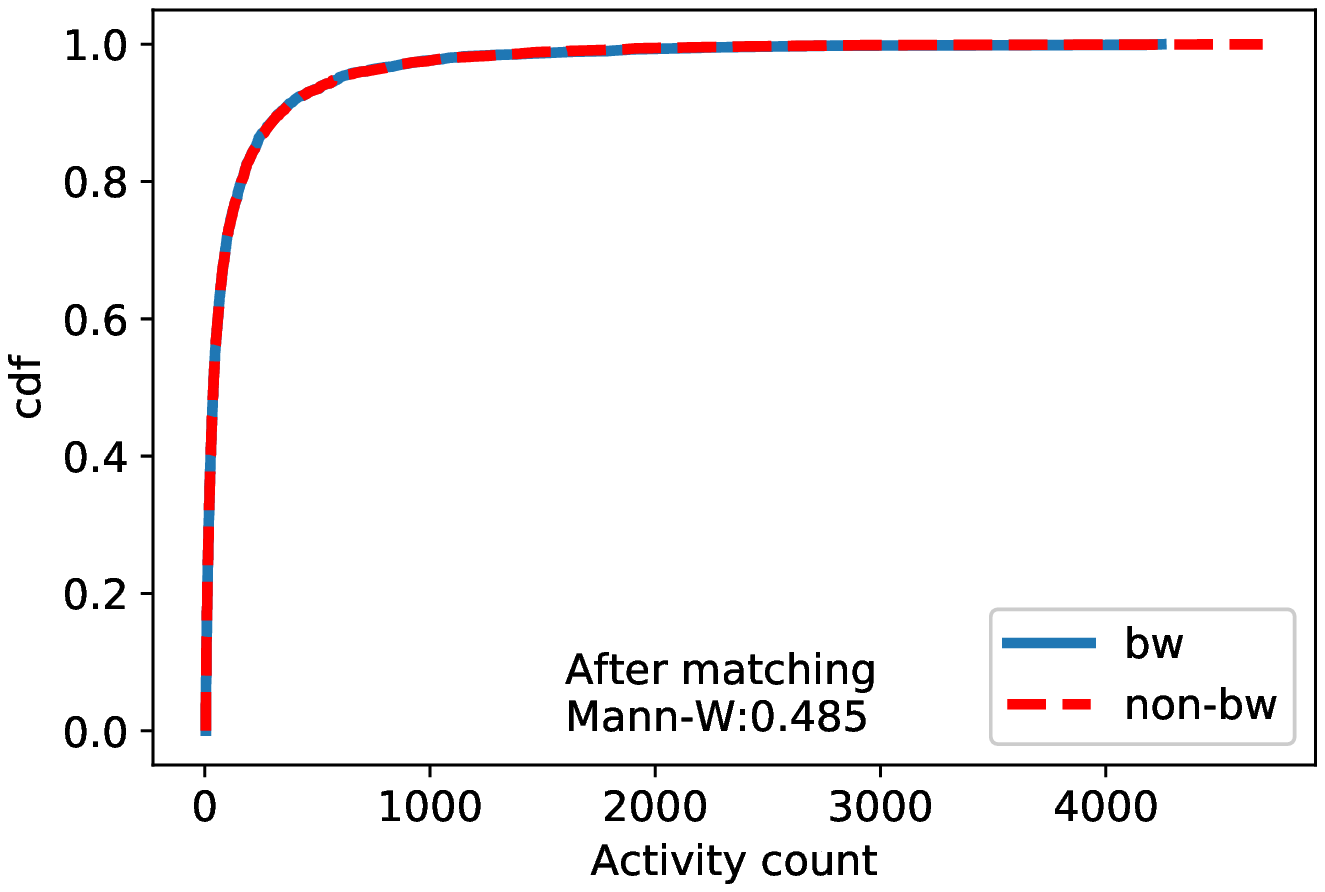}
      \caption{After, 2015-16}
    \end{subfigure}\hfil
    \begin{subfigure}[b]{.333\linewidth}
      \includegraphics[width=.99\textwidth]{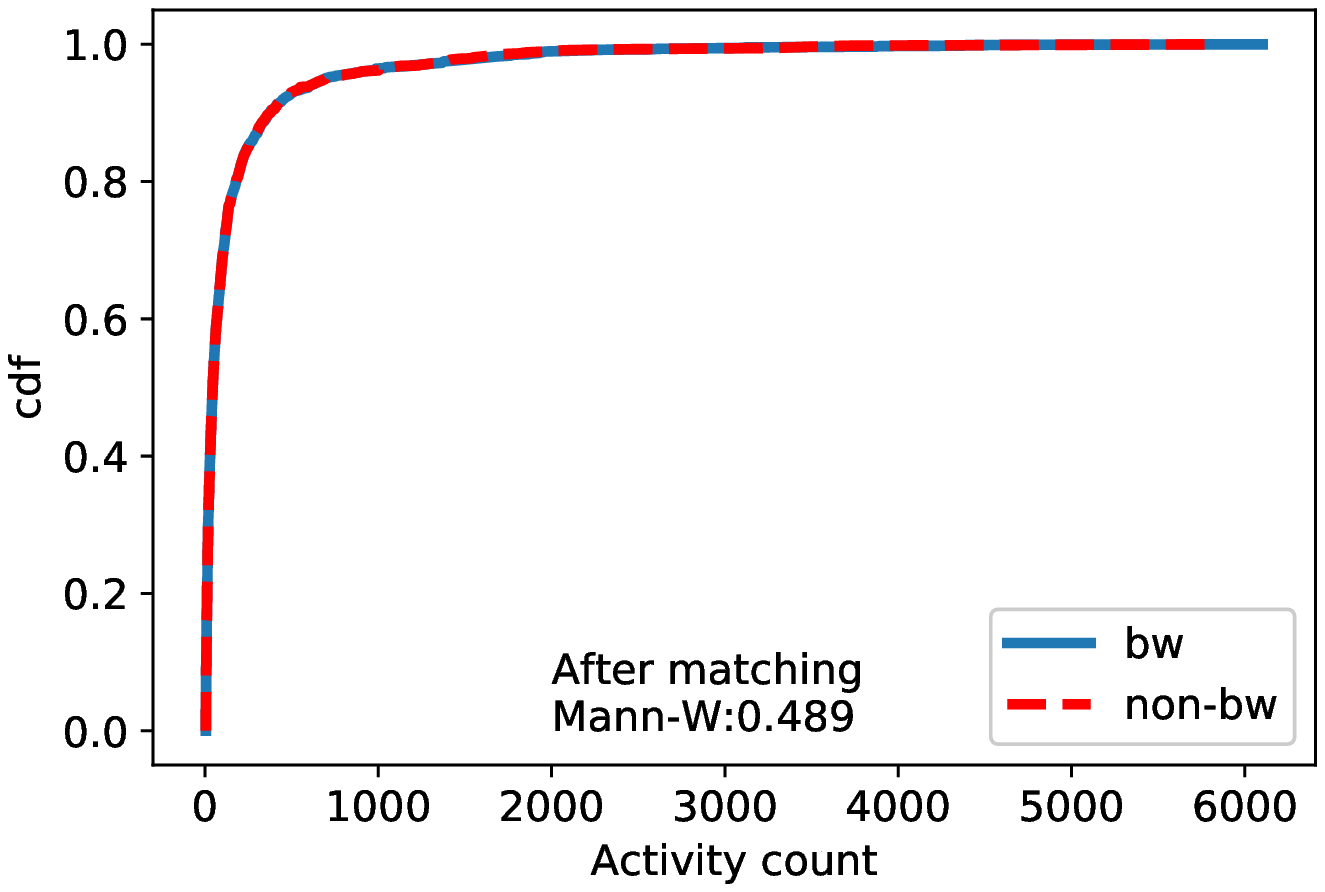}
      \caption{After, 2016-17}
    \end{subfigure}\hfil
    \begin{subfigure}[b]{.333\linewidth}
      
      \includegraphics[width=.99\textwidth]{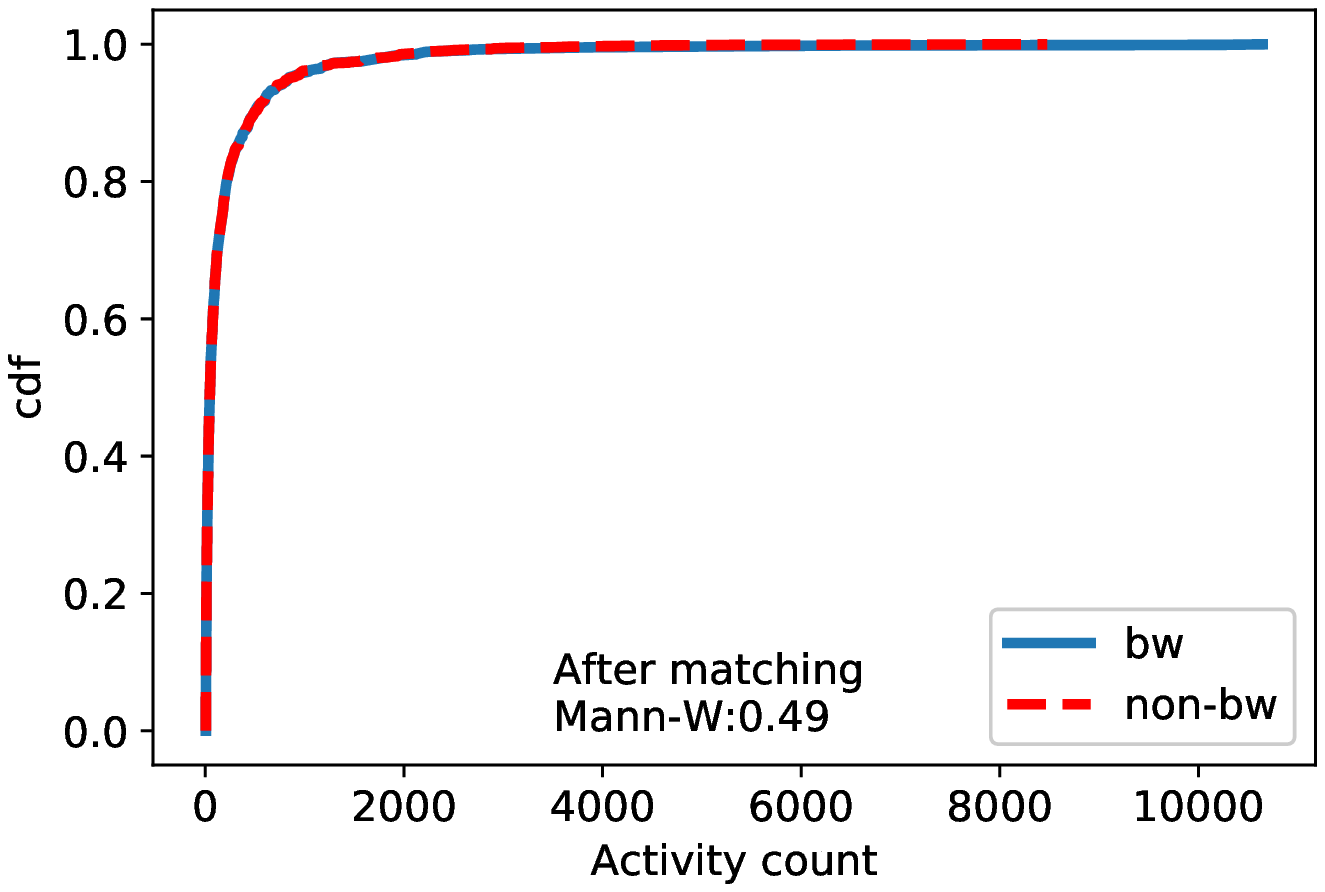}
      \caption{After, 2017-18}
    \end{subfigure}
    
\caption{Mann-whitney U test results before and after matching. Before matching: randomly select the same number of non-bandwagon users as the number of bandwagon users and run the test. After matching: run the test on the two matched user groups.}
\label{fig:uTest}
\end{figure}

\begin{figure}[!htbp] 
    \begin{subfigure}[b]{.5\linewidth}
      \includegraphics[width=.8\textwidth]{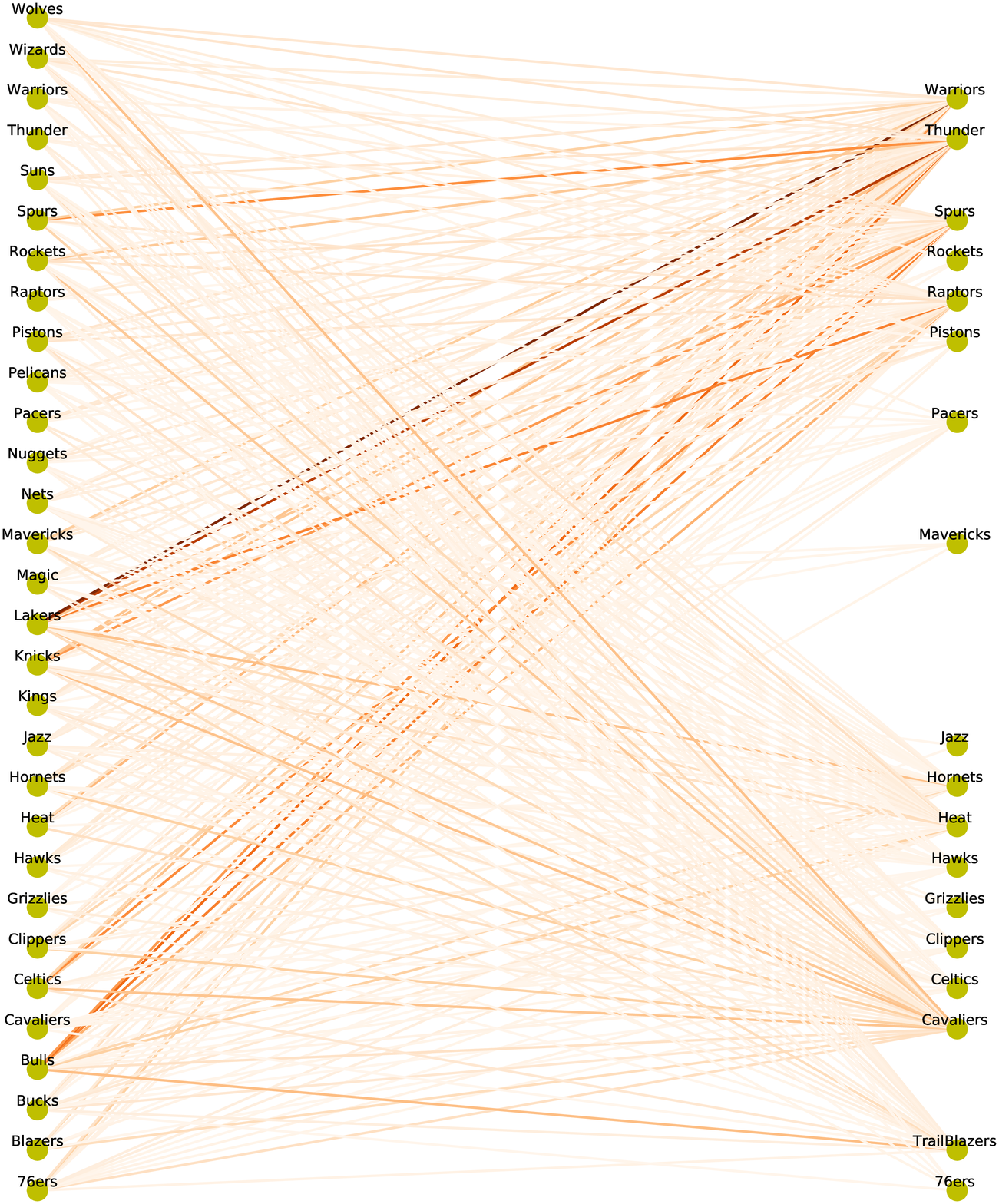}
      \caption{Stage 0}
    \end{subfigure}
    \begin{subfigure}[b]{.5\linewidth}
      \includegraphics[width=.8\textwidth]{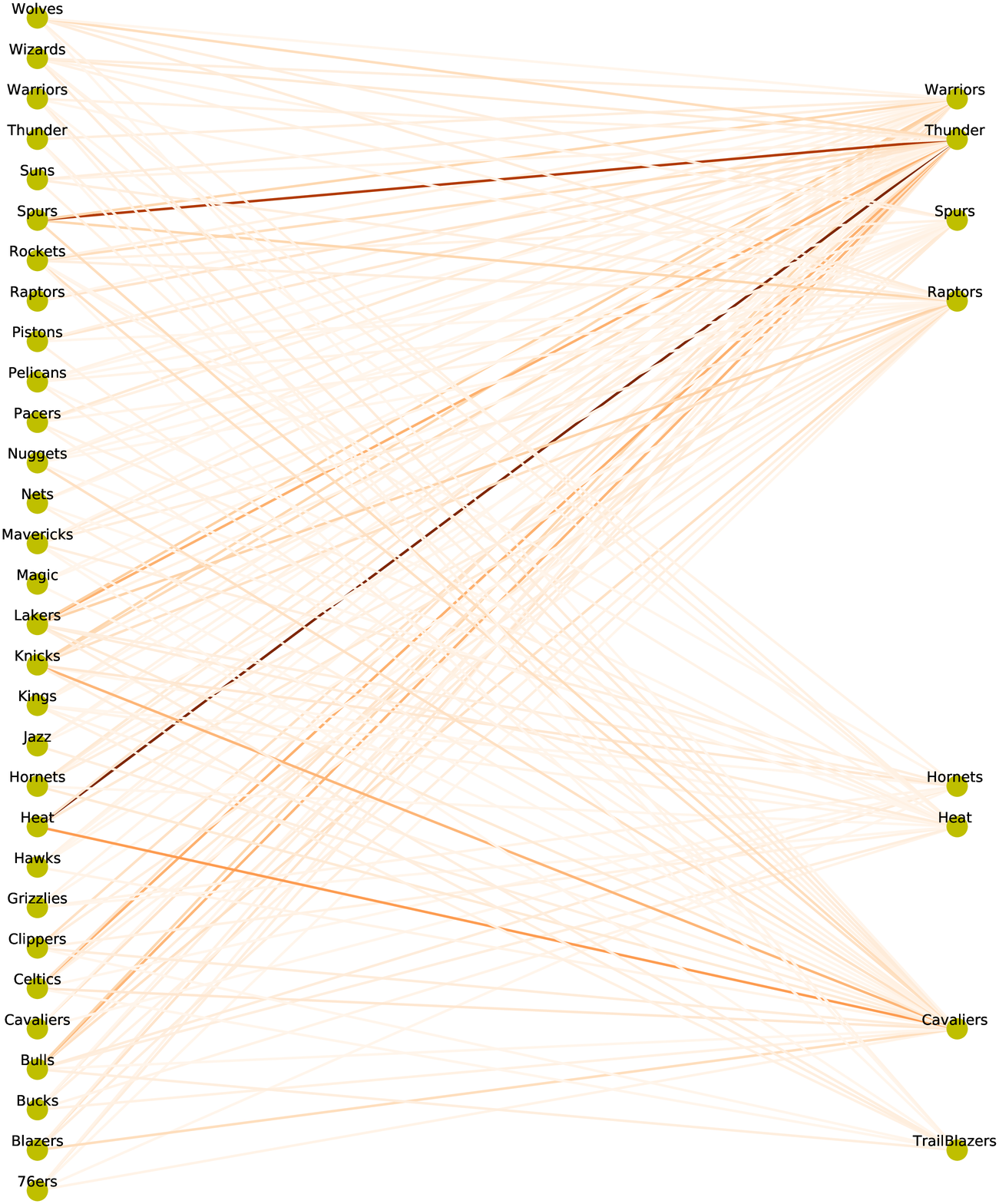}
      \caption{Stage 1}
    \end{subfigure}
    \begin{subfigure}[b]{.5\linewidth}
      \includegraphics[width=.8\textwidth]{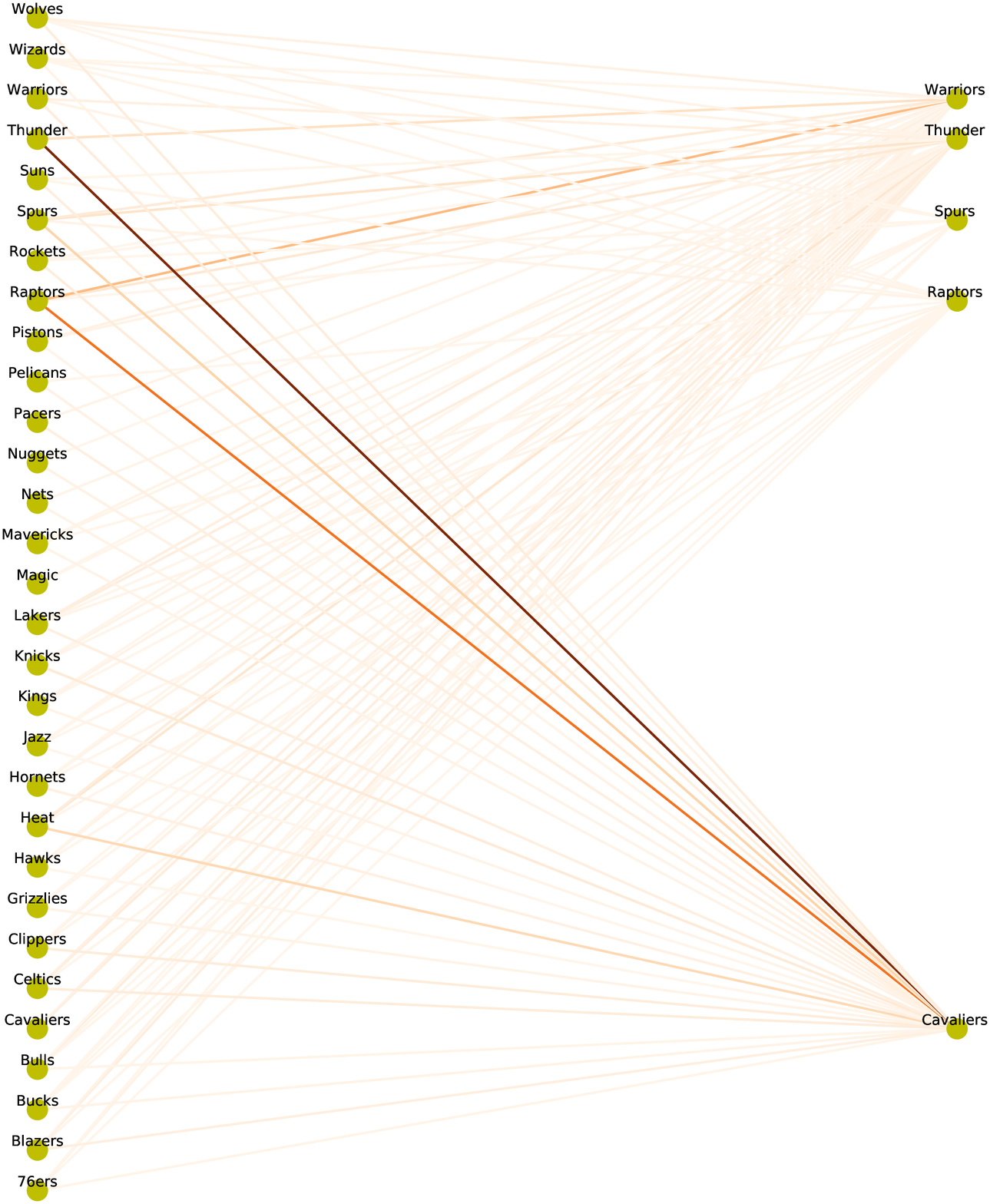}
      \caption{Stage 2}
    \end{subfigure}
    \medskip
    \begin{subfigure}[b]{.5\linewidth}
    
      \includegraphics[width=.8\textwidth]{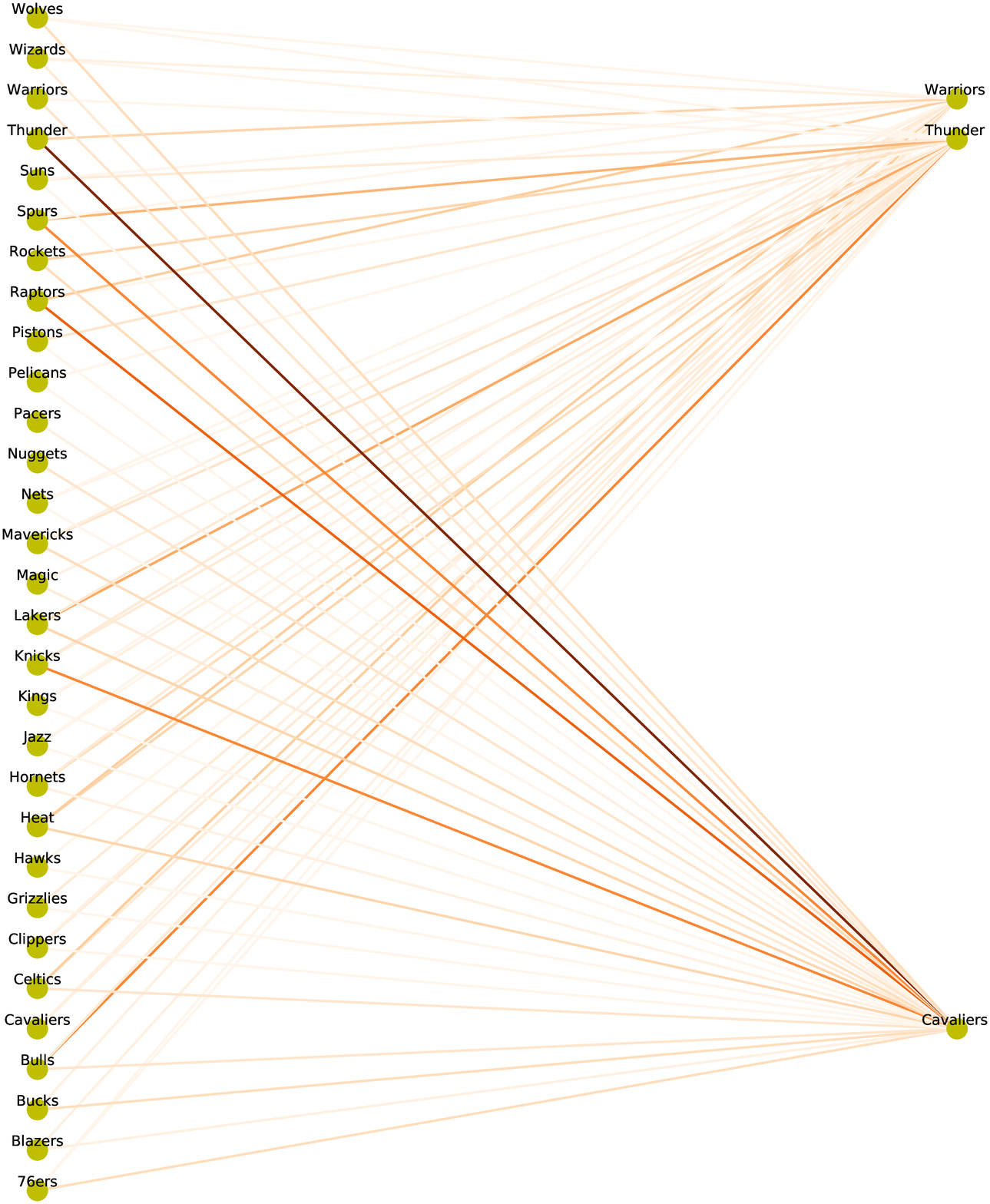}
      \caption{Stage 3}
    \end{subfigure}
    \begin{subfigure}[b]{.5\linewidth}
      \includegraphics[width=.8\textwidth]{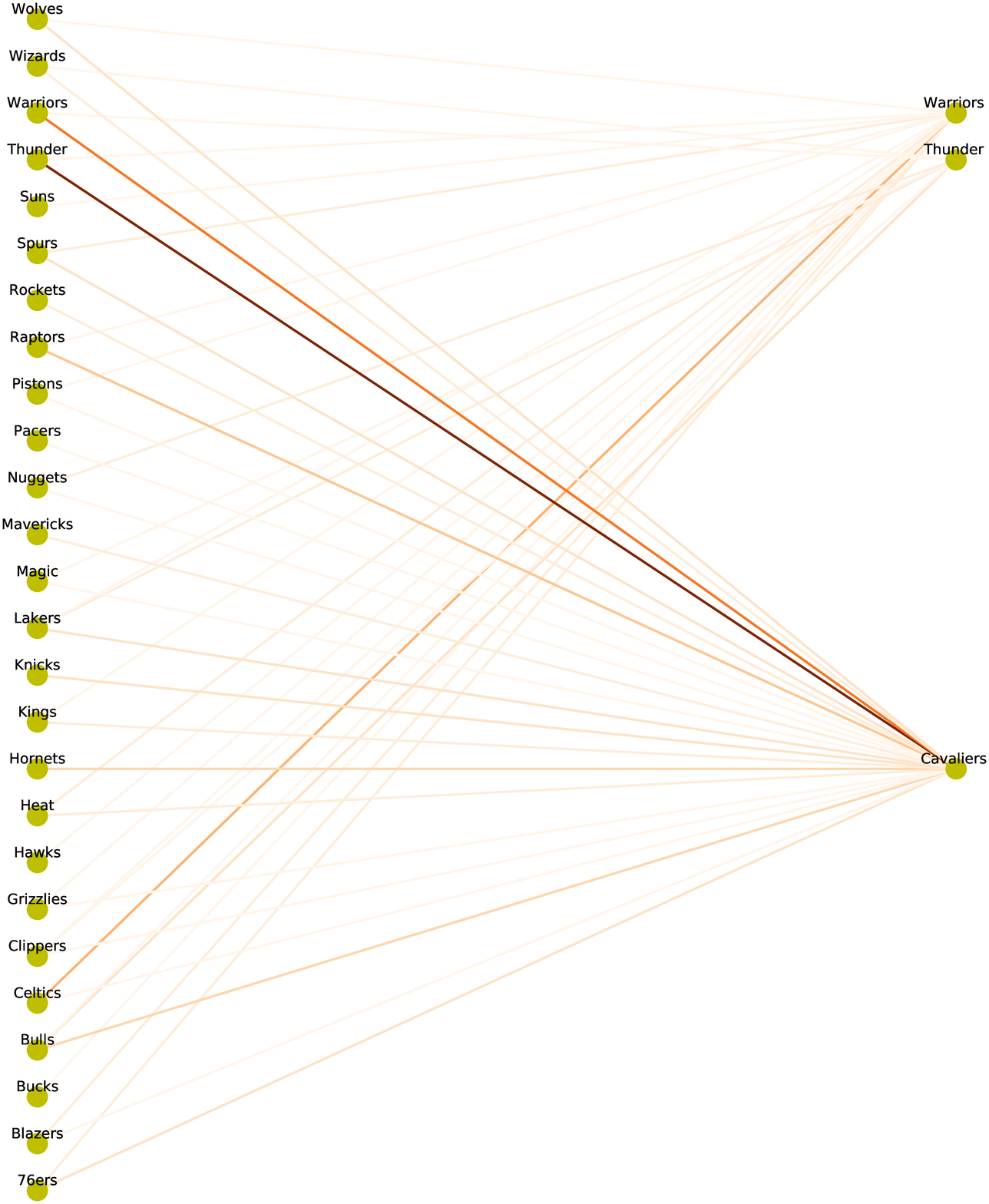}
      \caption{Stage 4}
    \end{subfigure}
\caption{Bandwagon fan flow in different stages in season 2015-16: From source team (team A, left nodes) to target team (team B, right nodes).}
\label{fig:flows16}
\end{figure}

Fig.~\ref{fig:flows1617} shows the general bandwagon fan flow of seasons 2015-16 and 2016-17, which validate Observation 1 in Section 4. Fig.~\ref{fig:flows16} is a stage-by-stage bandwagon fan flow. As shown in these flow graphs, bandwagon users keep leaving their original teams and going to stronger ones. There are fewer bandwagon users in later stages than that of the early stages.  In stages 3 and 4, only three teams remain as bandwagon target teams. Fig.~\ref{fig:uTest} shows the results of Mann-Whitney U test before and after user matching. 


\end{document}